# Modélisation multi-niveaux dans AA4MM

B. Camus[a,b]    J. Siebert[c]    C. Bourjot[a,b]    V. Chevrier[a,b]

Benjamin.Camus@loria.fr    Julien.Siebert@enib.fr    Christine.Bourjot@loria.fr    Vincent.Chevrier@loria.fr

a Université de Lorraine, LORIA, UMR 7506, Vandœuvre les Nancy, F-54506 France
b Inria, Villers-Lès-Nancy, F-54600 France
c Lab-STICC, Ecole Nationale d'Ingénieurs de Brest, France

*Résumé*

*Dans cet article, nous proposons de représenter un phénomène multi-niveaux sous la forme de plusieurs modèles en interaction. Cette vision structure la réflexion en rendant explicites les niveaux de représentation et leurs relations. Pour répondre aux défis de cohérence, causalité et coordination entre ces modèles, nous nous appuyons sur le méta-modèle AA4MM dédié à ce type de représentation. Un phénomène de flocking permet d'illustrer notre propos et de monter l'intérêt de cette approche.*

*Cette démarche pose des bases conceptuelles relativement simples pour aborder une question encore largement ouverte de la modélisation des systèmes complexes*

**Mots-clés** : *modélisation et simulation multi-niveaux ; émergence ; méta-modèle*

*Abstract*

*In this article, we propose to represent a multi-level phenomenon as a set of interacting models. This perspective makes the levels of representation and their relationships explicit. To deal with coherence, causality and coordination issues between models, we rely on AA4MM, a metamodel dedicated to such a representation. We illustrate our proposal and we show the interest of our approach on a flocking phenomenon.*

## 1. Introduction

Lorsque l'on aborde la modélisation de phénomènes collectifs, que ce soit en sciences humaines et sociales ou en biologie, il est souvent nécessaire de représenter le système selon plusieurs niveaux (Troitzsch, 1996).

Par exemple en biologie (Sloot, 2010) (Uhrmacher, 2009) (Ballet, 2012), le système à modéliser fait "naturellement" intervenir plusieurs niveaux ; tels celui des molécules, membranes, cellules, groupes de cellules, etc. ; qu'il est nécessaire de représenter et ainsi pouvoir étudier les liens entre ces niveaux.

Les systèmes multi-agents font partie des approches intégrant à la fois les aspects individuels et collectifs d'un système. Ils ont ainsi permis la modélisation de phénomènes où des individus en interactions locales font émerger un comportement collectif et permettent de répondre à des questions quant aux relations entre comportements individuels et propriétés collectives.

De manière plus générale, cette question s'inscrit dans l'étude des systèmes complexes. Ainsi dans (Chavalarias, 2008), les auteurs écrivent : "*La complexité des systèmes naturels et sociaux provient de l'existence de plusieurs niveaux d'organisation correspondant à différentes échelles spatio-temporelles. L'un des principaux défis de la science des systèmes complexes consiste à développer … des mé-*

*thodes de modélisation capables de saisir toutes les dynamiques d'un système par l'intégration de ses activités à de nombreux niveaux, souvent organisés hiérarchiquement.*"

Cet article tente d'apporter une première réponse à ce défi.

Après avoir situé la problématique de la modélisation multi-niveaux, nous proposons de représenter un phénomène multi-niveaux sous la forme de plusieurs modèles (ici on se limitera aux niveaux micro et macro) en interaction (partie 2). Cette vision structure la réflexion en rendant explicites les niveaux de représentation et leurs relations. Après avoir décrit un exemple de phénomène multi-niveaux simple, le flocking (partie 3), nous nous appuyons (partie 4) sur les concepts du méta-modèle AA4MM (*Agent et Artefact pour la Multi-Modélisatio*n) dédié à ce type de représentation pour réaliser une preuve de concepts autour de l'exemple choisi (partie 5).

## 2. La modélisation multi-niveaux

L'étude des phénomènes collectifs nécessite donc au moins deux niveaux de représentations (individuel et collectif). Les SMA offrent un paradigme intéressant car on a deux niveaux de discours qui sont le niveau micro (l'agent) et le niveau macro (le système). Cependant seul le niveau micro y est traditionnellement explicitement représenté, le niveau macro se contente le plus souvent d'être observé mais n'est pas défini en tant que tel. Par conséquent, ce niveau n'existe que par la présence d'un observateur extérieur au système et ne prend pas place dans un processus de modélisation. C'est ce que (Quijano, 2010) décrivent comme une approche "*mono-niveau dans la conception et bi-niveaux dans l'analyse des comportements produits*".

Il existe toutefois un certain nombre de travaux qui vont au delà de cette approche.

(David, 2009, David 2010, David 2011) proposent une démarche pour réifier les propriétés émergentes. Une première étape (processus d'introspection) consiste en la détection des propriétés émergentes. Cette détection s'appuie sur des connaissances du phénomène étudié. Ensuite ces propriétés sont réifiées (si besoin) sous la forme d'un agent émergent qui sera doté de propriétés et comportements. Enfin, il est possible de représenter l'influence de la propriété émergente sur le système sous la forme d'un élément d'interposition qui pourra modifier la perception et/ou l'influence des agents au niveau micro.

(Quijano, 2010) proposent une première analyse des manières dont peuvent être couplées des organisations multi-agents multi-niveaux à partir de différentes expériences de modélisation. Ils distinguent trois catégories de systèmes selon le type de couplage qui est réalisé entre les modèles associés à ces niveaux. Ils retiennent qu'il faut pouvoir intégrer des modèles existants aux formalismes potentiellement hétérogènes, ainsi que de pouvoir détecter, réifier et détruire dynamiquement les structures émergentes.

Traounez (Traounez, 2005) analyse le système au niveau micro pour détecter puis réifier des propriétés émergentes (en l'occurrence des tourbillons). De plus, il utilise une structure de donnée récursive pour représenter leur environnement à des échelles spatiales différentes. Cette représentation peut se rapprocher de celle de (Marilleau, 2008). Néanmoins, ces représentations correspondent à un changement d'échelle (plus ou moins précise) mais pas un changement de niveau de description.

Ces travaux considèrent tous que pour étudier un phénomène complexe à différents niveaux de description, il est nécessaire de représenter explicitement chacun des niveaux, de détecter puis réifier (Bonabeau, 1997) (si besoin) les propriétés collectives, et enfin, il doit être possible de contraindre le niveau micro à partir de considérations du niveau macro.

(David, 2010) (Traounez, 2005) proposent différents concepts mais ils sont pensés comme intégrés dans un tout. Ils n'abordent qu'insuffisamment les différents couplages possibles entre les niveaux (tels qu'évoqués dans (Quijano, 2010)). Notamment, les questions de comment représenter un phénomène à partir de modèles différents, de comment formaliser les couplages entre modèles qui correspondent aux relations entre niveaux de descriptions sont encore ouvertes.

(Bourgine, 2008) propose un cadre relativement général qui apporte un début de réponse à ces questions. Cette vision fait intervenir explicitement les modèles utilisés pour chaque niveau, ainsi que les relations entre ces niveaux. Le principe général est résumé par le schéma en figure 1.

L'idée sous-jacente est la suivante : on peut décrire un phénomène à un niveau macro si l'on est capable de caractériser ce phénomène par un ensemble d'informations (noté Y) et que l'on peut décrire l'évolution de Y temporellement sans avoir à faire référence au niveau micro. Cette vision exprime les différents éléments qui interviennent lors de la modélisation multi-niveaux et la manière dont ils s'articulent :

- X (resp. Y) correspond à l'ensemble des informations représentant le phénomène au niveau micro (resp. macro).

- f (resp g) est une fonction qui décrit la dynamique du phénomène au niveau micro (resp. macro).

- e est une fonction d'interprétation (émergence) permettant le passage du niveau de description micro à celui macro. Elle correspond aux méta-connaissances de (David, 2010).

- i correspond à l'immergence, c'est à dire, à l'influence du niveau macro sur le niveau micro (ce que (David, 2010) appelle fonction d'interposition). Cette influence modifie le comportement au niveau micro.

Si cette vision permet de situer et mettre en relation chacun des éléments, elle n'indique rien quant à leurs expressions (Quels formalismes pour exprimer e, i f ou g? Quelles correspondances entre t et t' ? Etc.).

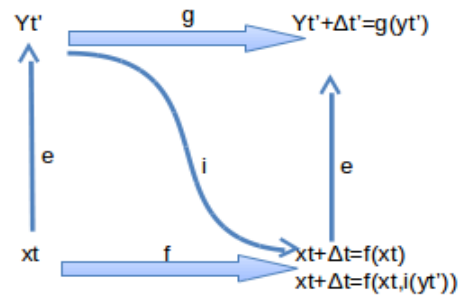

Figure 1 : Relations (d'après (Bourgine, 2008)) entre niveaux de description et évolution temporelle du phénomène.

Nous proposons de faire un pas vers une clarification de ce principe en envisageant chaque niveau comme un modèle et les relations entre eux comme des interactions.

La partie suivante présente un exemple jouet de modélisation multi-niveaux faisant intervenir les différents éléments évoqués par (Bourgine, 2008). Nous décrivons ensuite le métamodèle AA4MM qui nous permet de décrire cette représentation comme une société de modèles interagissant ; puis de l'implanter et la simuler.

## 3. Exemple de modélisation multi-niveaux

Cet exemple volontairement simple servira à illustrer notre proposition de modélisation multi-niveaux et à montrer comment s'instancient les différents concepts évoqués.

### 3.1. Les deux niveaux de modélisation du flocking

Nous prenons comme exemple les nuées d'oiseaux. Nous considérons deux niveaux de description : celui des oiseaux (microscopique) et celui des nuées (macroscopique).

La modélisation au niveau microscopique reprend le modèle multi-agent développé dans (Wilensky, 1998). L'état d'un oiseau est défini par un identifiant, une position et une orientation. Son comportement se déduit de ses comportements de cohésion, alignement et séparation vis à vis des autres oiseaux. La figure 2 visualise un état du niveau micro.

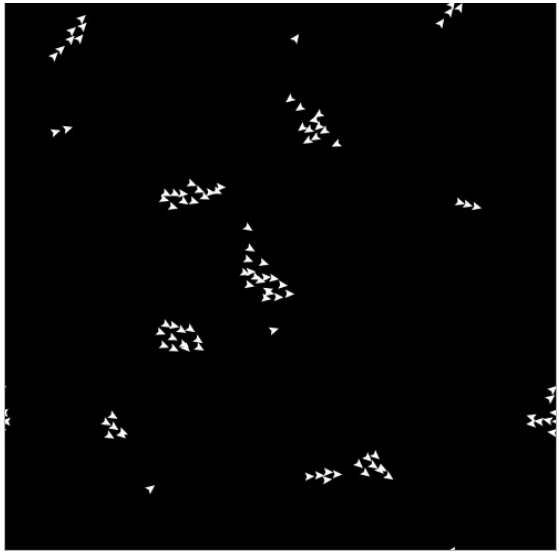

Figure 2: vue du niveau micro.

Au niveau macroscopique, une nuée est modélisée par un agent caractérisé par une position, une orientation et une taille. Nous supposons de manière arbitraire que le comportement d'une nuée se déduit de ses comportements de cohésion, alignement et séparation vis à vis des autres nuées.

Ce modèle macroscopique est une adaptation du modèle de (Wilensky, 1998) qui prend notamment en compte la taille de l'agent. Par simplification, il ne gère pas les fusions ni les séparations des nuées. La figure 3 visualise un état du niveau macro.

### 3.2. Les influences mutuelles

L'influence du niveau micro sur le niveau macro est fonction de l'interprétation que nous ferons d'une nuée (Cavagna, 2008) : si plusieurs oiseaux sont suffisamment rapprochés et possèdent des orientations proches, ils constitueront une nuée évoluant dans un déplacement commun.

Après avoir détecté les nuées, celles-ci seront réifiées dans le modèle macro par un agent situé au centre de gravité de la nuée et orienté selon la direction moyenne des oiseaux qui la composent, la taille est fonction de la dispersion. Ainsi, au fur et à mesure que les nuées apparaissent ou disparaissent, on met à jour le modèle au niveau macroscopique.

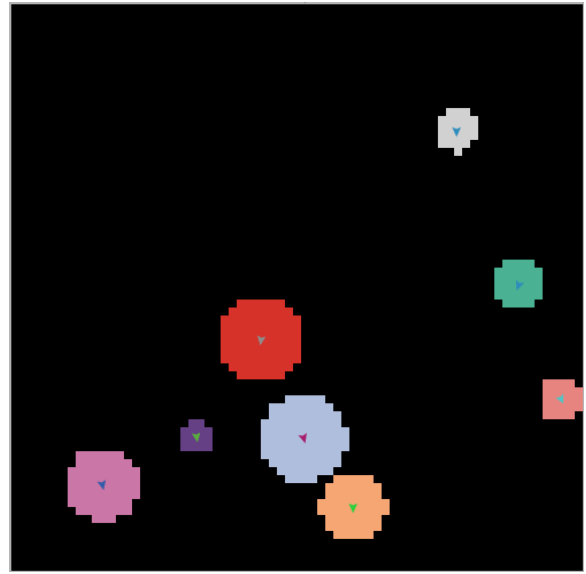

Figure 3: vue du niveau macro.

Nous supposons que l'influence du niveau macro sur le niveau micro est la suivante : tous les oiseaux appartenant à une nuée se déplacent de la même manière et ont la même orientation. Le modèle macroscopique calcule le déplacement de chaque nuée (vecteur de translation $\vec{v}$). Nous utilisons ce vecteur de déplacement global pour déplacer, au niveau micro chacun des oiseaux appartenant à la nuée. On remarquera qu'il faut donc étendre le modèle de (Wilensky, 1998) pour prendre en compte ce deuxième mode de déplacement au niveau microscopique.

Enfin, nous supposons également dans un premier temps que les dynamiques évoluent à la même vitesse.

La correspondance entre ce multi-modèle et le schéma inspiré de (Bourgine, 2008) s'effectue comme suit :

- X correspond aux caractéristiques des oiseaux,
- Y correspond aux caractéristiques des nuées,
- f est le modèle de déplacement des oiseaux,
- g est le modèle de déplacement des nuées,
- e est le mécanisme de détection des nuées,
- i est l'influence des nuées sur le comportement d'un oiseau.

## 4. Instanciation avec AA4MM

Nous proposons de représenter ce modèle multi-niveaux sous la forme de modèles (ici on se limitera aux niveaux micro et macro) en interaction (émergence et immergence). Pour cela, nous nous appuyons sur AA4MM, un méta-modèle dédié à ce type de représentation.

### 4.1. Le méta-modèle AA4MM

AA4MM (Siebert, 2011) modélise un phénomène complexe comme un ensemble de modèles en interaction (ce que nous appellerons par la suite un multi-modèle). Il s'inscrit dans le paradigme multi-agent : à tout modèle est associé un agent et les interactions sont supportées par des artefacts (Ricci, 2007).

On retrouve ce type d'approche également dans Reiscop (Desmeulles, 2009), Ioda (Kubera, 2011) ou encore Geamas (Marcenac, 1998). Les dynamiques du phénomène (micro/macro) sont représentées par des modèles différents. La dynamique globale du phénomène est simulée grâce à l'interaction des modèles. L'originalité vis à vis d'autres approches de multi-modélisation est d'envisager les interactions de manière indirecte, supportées par un environnement, et ainsi de modéliser explicitement le partage d'informations entre modèles (couplage structurel).

L'intérêt d'un méta-modèle dans notre cas est de disposer d'un cadre conceptuel dans lequel nous pourrons décrire les niveaux de représentation et la manière dont ils interagissent, c'est à dire modéliser un système sous la forme de dynamiques à différents niveaux ainsi que de leurs relations (influences).

D'un point de vue méthodologique, cette vision oblige à expliciter les choix de modélisation relatifs à chacun des modèles et à leurs interactions dès la phase de conception du multi-modèle et nous contraint sur la manière de les exprimer. La contrepartie est que AA4MM possède des spécifications opérationnelles et des algorithmes prouvés concernant la cohérence temporelle entre modèles qui permettent d'implanter le multi-modèle et de le simuler en ne devant coder qu'un nombre restreint de fonctions spécifiques à l'application visée. En effet, les spécifications opérationnelles permettent de disposer d'un intergiciel.

Cette approche a donné lieu à des preuves de concepts avec Netlogo (Siebert, 2010) et une application dans le cadre des réseaux mobiles ad-hoc (Leclerc, 2010) notamment en réutilisant des modèles existants et hétérogènes.

### 4.2. Représentation multi-niveaux avec AA4MM

Nous allons maintenant expliciter la représentation multi-niveaux à l'aide des composants de AA4MM. Il repose sur trois concepts à partir desquels il est possible de décrire un multi-modèle :

1. le *m-agent* contrôle un modèle et prend en charge les aspects dynamiques des interactions de ce modèle avec les autres modèles (figure 4a),
2. chacune de ces interactions (entre m-agents) est réifiée par un *artéfact de couplage* (figure 4b),
3. enfin l'*artéfact d'interface* réifie les interactions entre un m-agent et son modèle (figure 4c).

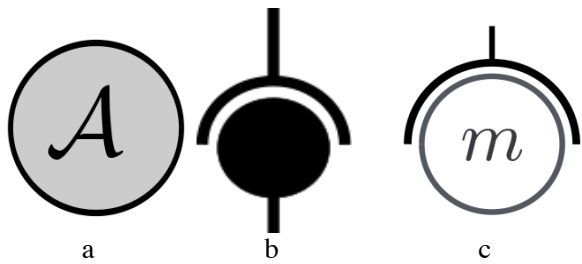

Figure 4 : symboles des composants du méta-modèle AA4MM (a) m-agent, (b) artéfact de couplage, (c) artéfact d'interface et son modèle *m*.

Le comportement d'un m-agent correspond au cycle :

- lecture des informations en provenance des artefacts de couplage,
- mise-à-jour et exécution de son modèle via l'artéfact d'interface,
- écriture d'informations à destination des artefacts de couplage.

Un artéfact de couplage propose des opérations d'écriture (un m-agent écrit les informations à transmettre), de transformations (changement d'unités par exemple), et de lecture (un m-agent lit des informations).

Un artéfact d'interface autorise les opérations suivantes sur le modèle : initialisation des données, mise à jour des caractéristiques individuelles, exécution d'un pas de simulation.

### 4.3. Démarche de modélisation d'un phénomène multi-niveaux

La création d'un multi-modèle via AA4MM se déroule en plusieurs étapes que nous déclinons ci-après dans le cas de la modélisation multi-niveaux.

La première étape consiste à définir le graphe d'interaction entre les modèles, c'est à dire, quelles informations sont échangées.

Dans notre cas, ce graphe se déduit assez simplement. Chaque niveau de représentation est associé à un modèle (noté *m* pour le niveau micro et *M* pour le niveau macro). Les interactions sont les relations d'émergence (notée *e*) de micro vers macro et d'immergence (notée *i*) de macro vers micro (voir figure 5).

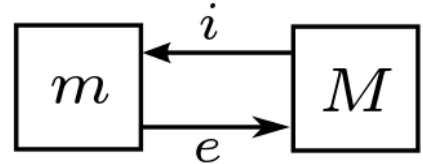

Figure 5 : Graphe des relations entre modèles.

La seconde étape consiste à exprimer ce graphe à l'aide des concepts de AA4MM. Ainsi à chaque modèle, et donc à chaque niveau, correspond un m-agent et un artefact d'interface. Le m-agent $A_m$ contrôle (gère) le modèle du niveau micro et le m-agent $A_M$ contrôle le modèle du niveau macro. A chaque relation entre modèles correspond un artefact de couplage entre leurs m-agents correspondants (cf. figure 6).

La mise en œuvre de ce diagramme nécessite des extensions de AA4MM par rapport à ses spécifications initiales. En détaillant notre exemple, nous soulignons les manques et indiquons[1] la manière dont nous y répondons ci-après.

Concernant les artéfacts d'interface, nous réutiliserons ceux existant pour Netlogo dans la bibliothèque AA4MM.

Dans une démarche bottom-up, initialement le niveau micro comporte un certain nombre d'oiseaux créés aléatoirement. Le m-agent $A_m$ fait exécuter un pas de simulation, récupère les identifiants, positions et orientations des oiseaux et les écrit dans l'artefact de couplage *e*.

L'artefact de couplage *e* reçoit donc du m-agent $A_m$ la liste des positions et orientation de chacun des oiseaux et les interprète pour construire des nuées. Ceci est réalisé par un algorithme d'identification de cluster avec deux paramètres : un seuil de proximité et un seuil d'orientation. Ces informations (liste de nuées avec centre de gravité, orientation et taille) sont alors disponibles pour le m-agent $A_M$.

---

[1] La place disponible dans cet article ne nous permet pas de fournir l'intégralité des spécifications. Nous nous contentons d'indiquer les modifications qualitativement.

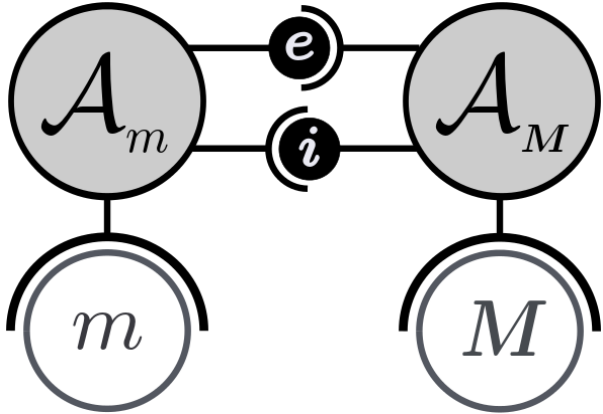

Figure 6 : Diagramme AA4MM de la représentation multi-niveaux.

Dans sa conception originelle, un artéfact de couplage conserve le cardinal des informations qui lui sont fournies lorsqu'il les transmet. La relation d'émergence oblige à concevoir un nouveau type d'artéfact (dit *d'interprétation*) qui dans le cas de la relation d'émergence réduit les informations (ici en taille). De plus, il se peut qu'il n'y ait aucune nuée.

Le m-agent $A_M$ lit les informations interprétées par l'artéfact de couplage $e$ ; met à jour les nuées répertoriées, ajoute les nouvelles et supprime celles disparues dans le simulateur ; fait exécuter un pas de simulation ; récupère les nouvelles positions des nuées ; calcule les vecteurs de déplacement des nuées puis écrit ces informations dans l'artéfact de couplage $i$.

On constate que le m-agent doit pouvoir ajouter ou supprimer des nuées dans le modèle, qu'il doit également traiter l'absence d'informations en provenance de l'artéfact d'émergence. Nous avons, en conséquence, adapté le comportement du m-agent et ses spécifications à ces besoins. L'artefact d'interface est étendu pour supporter ajout et suppression de nuées dans le modèle.

L'artéfact de couplage $i$ reçoit une liste de nuées accompagnées chacune de son vecteur de déplacement, transforme ces informations pour fournir une liste d'oiseaux accompagnées de leur vecteur déplacement et les rend disponibles au m-agent $A_m$.

Cet artéfact est également un artéfact d'interprétation car il y a augmentation de la quantité d'informations transmises.

Le comportement du m-agent $A_m$ consiste à lire les données en provenance de l'artéfact de couplage $i$, à faire exécuter un pas de simulation en tenant compte des deux modes de déplacement, à récupérer les identifiant, position et orientation de chacun des oiseaux, puis à écrire ceux-ci dans l'artéfact de couplage $e$. Comme il est possible qu'il n'y ait pas de nuée, ce m-agent doit être étendu pour traiter l'absence d'information en provenance de l'artéfact d'immergence.

## 5. Expérimentations

L'objectif des expérimentations est de montrer la faisabilité de notre proposition et d'en illustrer l'intérêt et les possibilités.

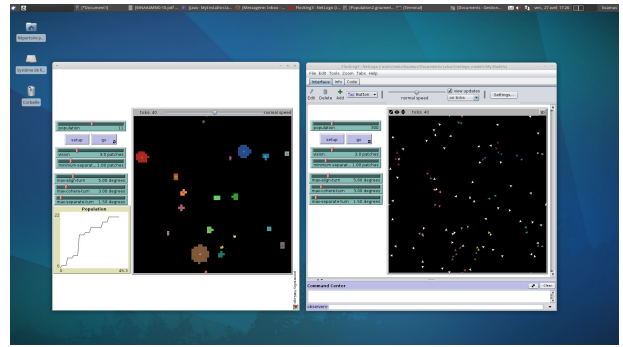

Figure 7 : exécution du multi-modèle.

Nous avons donc implanté l'exemple du flocking. La figure 7 correspond à une copie d'écran de la simulation multi-niveaux. Il faut noter qu'ici les modèles s'exécutent dans deux instances Netlogo (Wilensky, 1999) séparées.

Concernant l'intérêt de l'approche au niveau du "multi-modélisateur", nous montrons son pouvoir explicatif (cf 5.1), sa capacité à construire différentes déclinaisons du multi-modèle par changement de modèles (cf 5.2), de couplages (cf 5.3) ou de dynamique temporelle (cf 5.4) et à comparer ces déclinaisons (cf 5.5).

Les propriétés prouvées de AA4MM (cohérence, causalité et coordination entre modèles (Siebert, 2011)) facilitent le développement des différentes déclinaisons.

## 5.1. Niveaux micro-macro explicites

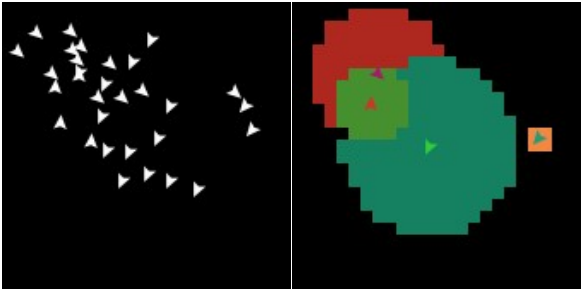

Figure 8 : croisement de plusieurs nuées (gauche niveau micro, droite niveau macro).

L'existence de deux niveaux de description distincts utilisant chacun un vocabulaire différent (Muller, 2004) permet de rendre compte de phénomènes, comme par exemple, exprimer explicitement le croisement entre plusieurs nuées (invisible au niveau micro mais visible au niveau macro, voir figure 8).

## 5.2. Influence du modèle macro sur le phénomène

Outre le modèle $M$ décrit en section 3, nous avons utilisé deux autres modèles (dérivés de $M$) pour le niveau macro : $M1$ utilise un fort facteur de séparation et de faibles facteurs de cohésion et alignement ; $M2$ utilise de forts facteurs de cohésion et alignement et un faible facteur de séparation.

## 5.3. Phénomène sans immergence

Nous souhaitons également étudier le comportement du phénomène sans immergence. L'implantation se déduit à partir de la figure 9 : les composants inutiles sont supprimés et les m-agents modifiés de la manière suivante : $A_M$ se contente de lire les informations en provenance de $e$, $A_m$ n'a plus besoin de lire les données en provenance de $i$ (lequel a disparu).

Cette déclinaison sera notée $m$ en figure 10.

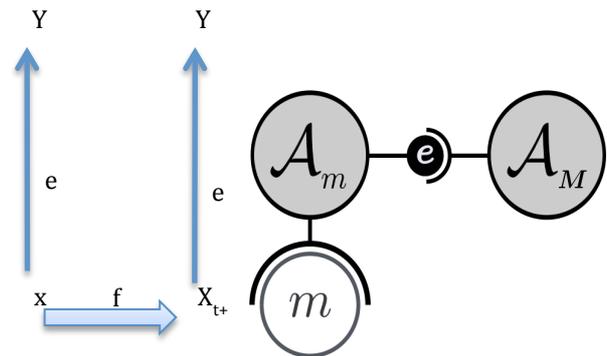

Figure 9 : couplage sans immergence, ni comportement macro.

## 5.4. Couplage avec des échelles temporelles différentes

Nous reprenons notre exemple tel qu'en section 3 mais en supposant maintenant des échelles de temps différentes : il y a quatre pas de simulation au niveau micro pour un au niveau macro (noté ici $M3$).

Nous revisitons l'interprétation que nous faisons de l'immergence : le déplacement macro est décomposé linéairement en quatre déplacements micro. L'artéfact de couplage $i$ transforme donc les données qu'il reçoit pour construire quatre listes associant à un identifiant un quart de vecteur déplacement. Au niveau de la cohérence temporelle, nous tirons parti de l'algorithme d'exécution prouvé de AA4MM qui gère les relations temporelles et ne modifions que $i$ et la correspondance temporelle au niveau de $A_{M3}$.

## 5.5. Comparaison des déclinaisons du multi-modèle

Dans notre exemple, l'évolution du nombre de de nuées dépend de plusieurs facteurs : le comportement individuel micro, et en cas d'immergence du niveau macro au travers des paramètres du comportement des nuées.

La figure 10 montre une comparaison de l'influence de ces différents facteurs. Chaque point d'une courbe correspond à 100 exécutions.

Dans une démarche de modélisation, ce type d'étude est important pour comprendre les relations entre niveaux. Notre approche permet de situer précisément quelle partie du multi-modèle est impliquée.

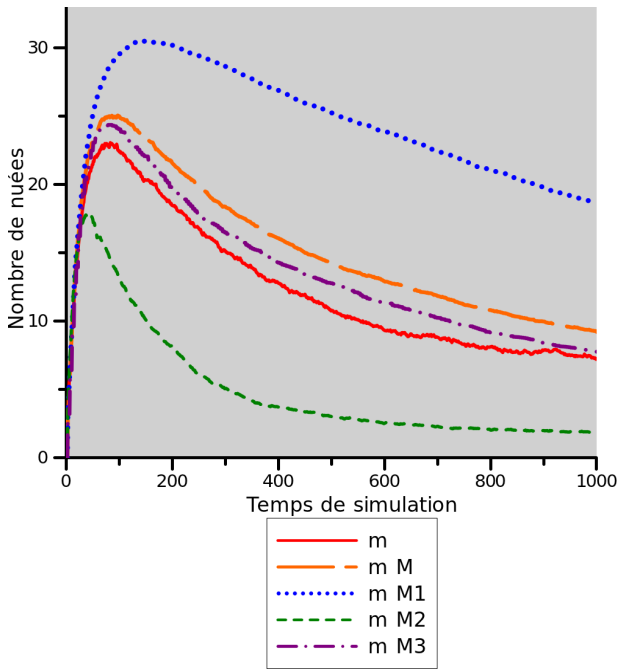

Figure 10 : comparaison du nombre de nuées selon l'influence de différents facteurs. Les notations m, M, M1, M2 et M3 font référence aux modèles présentés précédemment.

## 6. Conclusion

Dans cet article, nous avons apporté une première réponse au problème de la représentation multi-niveaux de phénomènes complexes.

Notre proposition, au niveau conceptuel, s'inspire de (Bourgine, 2008) qui envisage la représentation multi-niveaux comme un couplage entre plusieurs systèmes dynamiques, ce que nous exprimons de façon modulaire sous la forme de plusieurs modèles multi-agents en interaction. Pour répondre aux défis de cohérence, causalité et coordination entre ces modèles, nous nous appuyons sur le méta-modèle AA4MM dédié à ce type de représentation.

AA4MM nous offre de par son approche modulaire une flexibilité que nous pouvons exploiter dans le cadre d'une modélisation multi-niveaux. Cela permet facilement le changement de modèles, le changement d'échelle temporelle, le changement de couplage entre niveaux ; ainsi que la réutilisation de modèles existants.

La démarche sous-jacente propose de partir du problème de modélisation multi-niveaux, pour établir un graphe de relations (émergence/immergence) entre les deux niveaux micro/macro que nous traduisons sous la forme d'un diagramme AA4MM. Ce diagramme est alors implanté à partir de l'intergiciel de AA4MM.

Ce travail pose des bases conceptuelles relativement simples pour aborder une question encore largement ouverte de la modélisation des systèmes complexes. Nous envisageons d'étendre cette approche à des systèmes qui ne sont plus simplement bi-niveaux (micro-macro), et de la confronter à d'autres phénomènes dont les relations d'émergence et d'immergence prennent d'autres formes.

## Remerciements

## Références